\documentclass[aps,prb,groupedaddress]{revtex4}
\usepackage{graphicx}
\usepackage{dcolumn}
\usepackage{bm}
\begin{document}
\preprint{APS/123-QED}
\title{Chirality, charge and spin-density wave instabilities of a two-dimensional electron
gas in the presence of Rashba spin-orbit coupling}
\author{George E. Simion}
\author{Gabriele F. Giuliani}
\affiliation{Physics Department, Purdue University}

\date{\today}

\begin{abstract}
We show that a result equivalent to Overhauser's famous Hartree-Fock instability
theorem can be established for the case of a two-dimensional electron gas
in the presence of Rashba spin-obit coupling. In this case it is the spatially
homogeneous paramagnetic chiral ground state that is shown to be differentially
unstable with respect to a certain class of distortions of the spin-density-wave and
charge-density-wave type. The result holds for all densities. Basic properties of
these inhomogeneous states are analyzed.
\end{abstract}
\pacs{71.10.Ay, 71.10.Ca, 72.10.-d, 71.55.-i}
\keywords{two dimensional electron gas, many-body effects, spin-orbit coupling}

\maketitle

\section{Introduction}
Recent interest in the properties of the quasi-two dimensional
electron and hole devices in the presence of structural
(Rashba-Bychkov)\cite{Rashba1,Rashba2} or intrinsic
(Dresselhaus)\cite{Dress} spin orbit has brought to the fore the
problem of the interacting chiral electron liquid. It is therefore
important to revisit several of the fundamental notions of many-body
theory for this intriguing system. The purpose of present paper is
to begin a theoretical exploration of the relevance and special
properties of a class of spatially non homogeneous spontaneously
broken symmetry states of the electron liquid in the presence of
Rashba spin-orbit coupling in two dimensions. Specifically we will
focus our attention on spin density and charge density wave type
states henceforth referred to for simplicity as SDW and CDW. SDW and
CDW states, originally conceived by A. W.
Overhauser,\cite{Over1,Over2} are generally stabilized by the
electron-electron interaction and are characterized by spatial
oscillations of the spin density, the charge density, or both. In
the absence of spin-orbit coupling, one can begin to describe SDW
and CDW states by simply considering the electron number density for
both spin projections:
\begin{equation}
n_{\uparrow}=\frac{n}{2}+A \cos \left(\vec{Q} \cdot
\vec{R}+\frac{\phi}{2}\right)~,
\end{equation}
\begin{equation}
n_{\downarrow}=\frac{n}{2}+A \cos \left(\vec{Q} \cdot
\vec{R}-\frac{\phi}{2}\right)~.
\end{equation}
In these expressions the wave vector $\vec{Q}$ spans the Fermi surface, i.e. does
satisfy the condition $|\vec{Q}| \simeq 2k_F$. A CDW corresponds to $\phi=0$, while a
SDW state obtains for $\phi=\pi$. Mixed state are also possible: one such state is
beautifully realized
in chromium.\cite{AWO_Arrott_chromium,Corliss_et_al_on_chromium,Shull_Wilkinson_RMP_neutron_diffr_chromium}
As we will show, in the presence of linear Rashba spin orbit the corresponding distorted states
are characterized by a more complex spatial dependence of the number density, the spin density
and, where appropriate, a chiral density.
As a first step towards establishing the fundamental properties of SDW- and CDW-like states
in the presence of spin-orbit interaction we present here a generalization of the famous
Overhauser's Hartree-Fock (HF) instability theorem. The latter represents an important
exact result in many-body theory for it establishes that, within HF, the homogeneous paramagnetic
plane wave state does not represents a minimum of the energy for an otherwise uniform
electron gas for it can be always variationally bettered by a suitably constructed distorted chiral
SDW or CDW.\cite{comment_Hartree_term_cancellation}

The paper is structured as follows: In Section \ref{sec:SO} we
discuss the relevant aspects of the theory of a two dimensional non-interacting
electron gas in the presence of Rashba spin-obit coupling. Section \ref{RashbaHF}
briefly discusses useful notions of the electron-electron interaction within Hartree-Fock
approximation. Section \ref{proof1} is dedicated to the actual proof of the theorem and
contains the main results. Finally the last Section contains the conclusions while a number
of useful mathematical relations are derived in the two Appendices.


\section{Two dimensional electron gas in the presence of Rashba spin-orbit}
\label{sec:SO}
In the presence of linear Rashba spin-orbit coupling, the one-particle hamiltonian
can be written as follows:
\begin{equation}
\label{NIHRashba} \hat H_0=\frac{\vec{p}^2}{2m}+\alpha
\vec{p}\cdot\left(\vec{\sigma}\times \hat{z}\right)~,
\end{equation}
where $\hat{z}$ is the unit direction along the z-axis, the motion taking place
in the $x,y$ plane.

\begin{figure}[h]
\centerline{\includegraphics[width=4 in, height=3 in]{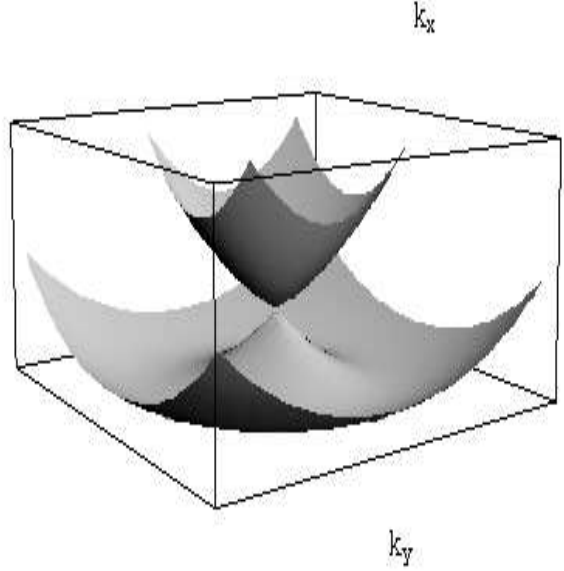}}
\caption{Non interacting energy spectrum in the presence of linear Rashba spin-orbit
coupling.} \label{SpaceE}
\end{figure}

The non interacting problem can be readily diagonalized to obtain the energy
spectrum and the eigenfunctions:
\begin{equation}
\label{en_nonint_k} E_{k,\mu}=\frac{\hbar^2 k^2}{2m}-\alpha \mu k~,
\end{equation}
and
\begin{equation}
\label{wf_nonint_p}
\psi_{\vec k,+}=\frac{1}{\sqrt{2}L}e^{i\vec{k} \cdot \vec{r}} \left(%
\begin{array}{c}
1 \\
-ie^{i\phi_{\vec{k}}} \\
\end{array}%
\right)~,
\end{equation}
\begin{equation}
\label{wf_nonint_m}
\psi_{\vec k,-}=\frac{1}{\sqrt{2}L}e^{i\vec{k} \cdot \vec{r}} \left(%
\begin{array}{c}
-ie^{-i\phi_{\vec{k}}}\\
1 \\
\end{array}%
\right)~,
\end{equation}
where $\mu= \pm$ labels a state's {\it chirality} and $\phi_{\vec{k}}$ is
the angle spanned by the $x$-axis and the two dimensional wave vector $\vec{k}$.
A schematic of the lower energy sector of the spectrum is plotted in Fig.~\ref{SpaceE}.

By making use of the states of Eqs.~(\ref{wf_nonint_p}) and
(\ref{wf_nonint_m}) as a basis for a second quantization
representation, the familiar fully interacting electron gas
hamiltonian reads:
\begin{widetext}
\begin{equation}
\label{HRashba_second} \hat
H\!=\sum\limits_{\vec{k},\mu}E_{\vec{k},\mu}\hat{b}_{\vec k, \mu}
\hat{b}^{\dag}_{\vec k, \mu} \!+\! \frac{1}{2L^2} \!
\sum\limits_{\vec q ,\vec{k}_1,\vec{k}_2 } \sum\limits_{
\mu_1,\mu_2, \mu_3, \mu_4 } \!\!\! v_q
\Phi_{\mu_1\mu_2 \mu_3 \mu_4 }^{\vec k _1, \vec k _2, \vec q }
\hat{b}^{\dag}_{\vec k _1 + \vec
q, \mu_1} \hat{b}^{\dag}_{\vec k _2 - \vec q, \mu_2} \hat{b}_{\vec k
_2,\mu_3} \hat{b}_{\vec k _1,\mu_4}~,
\end{equation}
\end{widetext}
where the phase factor
$\Phi_{\mu_1\mu_2 \mu_3 \mu_4 }^{\vec k _1, \vec k _2, \vec q }$
is defined as follows:
\begin{widetext}
\begin{eqnarray}
\label{phiSO} \Phi_ {\mu_1\mu_2 \mu_3 \mu_4 } ^{\vec k _1, \vec k
_2, \vec q } = i^{\frac{\mu_1+\mu_2-\mu_3-\mu_4}{2}}
e^{i\left(\frac{1-\mu_1}{2}\phi_{\vec k _1 + \vec q} +
\frac{1-\mu_2}{2}\phi_{\vec k _2 - \vec q} -
\frac{1-\mu_3}{2}\phi_{\vec k _2 } - \frac{1-\mu_4}{2} \phi_{\vec
k_2 } \right)} \times \nonumber \\
\frac{1}{4}\left[1+\mu_1\mu_4 e^{i\left(\phi_{\vec k _1} -
\phi_{\vec k _1 + \vec q}\right)}\right] \left[1+\mu_2\mu_3
e^{i\left(\phi_{\vec k _2} - \phi_{\vec k _2 - \vec
q}\right)}\right]~.
\end{eqnarray}
\end{widetext}

Uncorrelated many body wavefunctions for the system at hand can be
represented by Slater determinants constructed by occupying any
combinations of the chiral states Eqs.~(\ref{wf_nonint_p}) and
(\ref{wf_nonint_m}).

\section{Hartree-Fock theory of a two-dimensional electron liquid in the presence
of Rashba spin-orbit coupling}
\label{RashbaHF}

An accurate description of a realistic electronic system requires
that electron-electron interaction be taken into account. A first
step towards developing such a many-body theory is to investigate
the results of a mean-field approach. The main idea behind the mean
field procedure is to find an effective Hamiltonian which is
quadratic in the electron creation and annihilation operators and
can therefore be easily diagonalized. Within the HF theory, the
ground state is approximated by a single Slater determinant made out
of single particle wavefunctions, which in turn, are determined by
imposing the requirement that the expectation value of the
Hamiltonian over the Slater determinant be a minimum\cite{TheBook}.
Using these wavefunctions as our basis set, a standard Wick
decoupling procedure\cite{TheBook} allows us to determine the
effective HF potential. It can be easily proved that the
non-interacting chiral states are indeed among the solutions of the
corresponding HF equations. In this case, the HF potential is diagonal
in wave vectors and chiral indices:

\begin{equation}
V^{HF}_{\vec k\mu, \vec k'\mu'}=-\frac{\delta_{\vec k,\vec k'}
\delta_{\mu, \mu'}}{2L^2} \sum\limits_{\vec{\kappa} \nu} v_{\vec
k-\vec{\kappa}} \left[1+\mu \nu \cos(\phi_{\vec k}
-\phi_{\vec{\kappa}}) \right] ~.
\end{equation}

The corresponding HF eigenvalues are given by:
\begin{equation}
\label{EHF}
\epsilon_{\vec k \mu}=\frac{\hbar^2 k^2}{2m}-\alpha\mu k
-\frac{1}{2L^2} \sum\limits_{\vec{\kappa} \nu} v_{\vec
k-\vec{\kappa}} \left[1+\mu \nu \cos(\phi_{\vec k}
-\phi_{\vec{\kappa}}) \right]~.
\end{equation}

An evaluation of the Fermi energy of the two sub-bands leads to an
interesting problem. Since one band will, in general, acquire more
exchange energy than the other, this may result (in a first
iteration) in two different Fermi levels.  In order to equalize them
(for elementary stability reasons), electrons from one subband will
have to be moved to the other. This is the phenomenon of {\it
repopulation}.

The spatially homogeneous chiral states are just one of the possible
Hartree-Fock solutions. A detailed analysis of the possible
solutions corresponding to symmetric occupations in momentum space
can be done by systematically minimizing the total energy as a
function of spin orientation and generalized chirality of the
system\cite{ChesiGFG}. More general solutions correspond to
non-symmetric occupations of the single particle chiral states. The
problem has been studied and the corresponding very interesting
phase diagram has been explored\cite{ChesiGFG,ChesiGFG1}. As we will
presently discuss there also exists an interesting class of
spatially non-homogenous solutions to the problem.

\section{Proof of the instability theorem}
\label{proof1}

We will proceed by showing that it is always possible to lower the
energy of the homogeneous paramagnetic chiral ground state by
introducing a suitable real space distortion which is periodic with
wave vector $\vec {Q} = 2k_F \hat {x}$. The general approach follows
that of Fedders and Martin \cite{Martin} and is based on an Ansatz
which represents a generalization of that given by Giuliani and
Vignale for the case of the three dimensional electron
gas.\cite{TheBook}

Let us consider first the putative HF ground state of our many-body
system $\left|\Phi_S \right>$. A complete, and, as we shall see convenient,
description of this determinantal state can be achieved in terms of the
corresponding single-particle density matrix elements here given by:
\begin{equation}
\label{densitymatrix} \rho_{\alpha \beta}
=\left<\Phi_S\left|\hat{a}^{\dagger}_{\alpha}\hat{a}_{\beta}\right|\Phi_S\right>~,
\end{equation}
where here $\alpha$ and $\beta$ label the one-particle states which are
used to build the Slater determinant.

Now, within the space of Slater determinants, any slightly modification of the
state $\left|\Phi_S \right>$ can be described in terms of a corresponding
infinitesimal change of the single-particle density matrix elements.
Let us indicate such a change by $\delta\rho_{\alpha \beta}$. At this point the next
task consists in trying to evaluate the change of the total HF energy in terms
of these quantities.

Since $\left|\Phi_S \right>$ is a solution of the HF equations,\cite{comment_HFsols}
the first order variation in the energy must vanish so that the problem
at hand is reduced to determining the sign of the energy change to second order in the
$\delta\rho_{\alpha \beta}$'s. The relevant expression is therefore given by:\cite{TheBook}
\begin{widetext}
\begin{equation}
\label{DeltaEHF} \Delta^{(2)}E_{HF}= \frac{1}{2} \sum
\limits_{(\alpha,\beta)} \delta \rho_{\alpha \beta}
\frac{\epsilon_{\beta}-\epsilon_{\alpha}}{n_{\alpha}-n _{\beta}}
\delta \rho_{\beta \alpha} + \frac{1}{2} \sum
\limits_{(\alpha,\delta)} \sum \limits_{(\beta,\gamma)} \left(
v_{\alpha \beta \gamma \delta}-v_{\alpha \beta \delta \gamma}\right)
\delta \rho_{\alpha \delta} \delta \rho_{\delta \gamma}~,
\end{equation}
\end{widetext}
where the notation $(\alpha,\beta)$ means that only states situated
on opposite sides of the Fermi sea are considered in the summation.
In this formula, we indicate the Hartree-Fock eigenvalues as
$\epsilon_\alpha$ and the corresponding occupation numbers as
$n_\alpha$.

The next step in our procedure consists in constructing a Slater
determinant for which the HF energy is lower than
$\langle\Phi_S|H|\Phi_S\rangle$. In order to do, we follow
Overhauser's idea and choose the new one-particle states to be
suitable linear combinations of chiral plane waves states situated
near opposite points on the Fermi surface, the distortion being
limited to a very narrow strip. The width of this strip will play
the role of a variational parameter. We then carefully devise an
expression for the wave vector dependent amplitude of the coupling
between the plane waves and construct the corresponding Slater
determinant. In the last step, we calculate the change of the
Hartree-Fock energy due to this perturbation to leading order in the
distortion amplitude from Eq.~(\ref{DeltaEHF}).
\begin{figure}[h]
\centerline{\includegraphics[width=3.00in,height=2.00in]{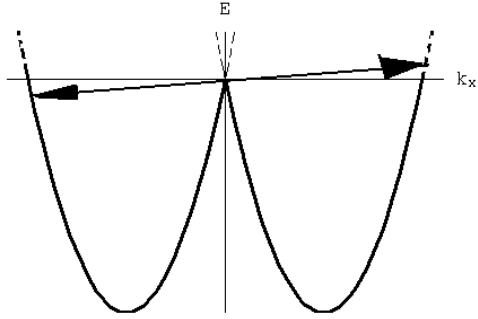}}
\caption{\label{Limitrs} Case of unit chirality when the states of the lower chiral
subband are occupied up to the lower threshold of the upper subband. The arrows
indicate the states that are coupled by the distortion along the $k_x$ axis.}
\end{figure}
\subsection{Instability for the case of chirality equal one}
\label{chi1}
Because it presents a formally simpler problem, the first case to be
treated is that in which only the lower subband is occupied while the upper one
is just about to be filled.\cite{comment_genChi1}
This situation is depicted in Fig.~\ref{Limitrs}.

Let us build the new trial wavefunctions as mixtures of the
wavefunctions corresponding to wave vector $\vec {k}$ with
those corresponding to wave vector $\vec {k}\pm \vec {Q}$, i.e.

\begin{equation}
\Psi_{\vec k} \simeq \psi _{\vec k , + } + A_{\vec k + \vec Q , \vec k}
\psi_{\vec k + \vec Q , + } + A_{\vec k - \vec Q, \vec k}
\psi_{\vec k - \vec Q , + }~.
\end{equation}
Here, as anticipated, $|\vec Q| = 2k_F^+$ as shown in Fig.
\ref{Limitrs}.

In evaluating the HF energy change, only the states situated on
opposite sides of the Fermi sea are relevant.  We will consider
$\left| {\vec {k}\pm \vec {Q}} \right|
> k_F^+ $ and $k < k_F $. Here, the occupation numbers are
$n_{\vec {k}} = \theta \left( {k_F - k} \right)$, while
the amplitude satisfies the condition: $A_{\vec {k}\pm \vec {Q},\vec
{k}} = A_{\vec {k},\vec {k}\pm \vec {Q}} $. The only non-zero matrix
elements of $\delta \hat{\rho}$ have the form:

\begin{equation}
\label{delt_rho_limit} \delta \rho _{\vec {k} + \frac{\vec
{Q}}{2},\vec {k} - \frac{\vec {Q}}{2}} = \delta \rho _{\vec {k} -
\frac{\vec {Q}}{2},\vec {k} + \frac{\vec {Q}}{2}} = \left( {n_{\vec
{k} + \frac{\vec {Q}}{2}} + n_{\vec {k} - \frac{\vec {Q}}{2}} }
\right)A_{\vec {k}}~.
\end{equation}

These wavefunctions indeed describe  SDW/CDW-like states. A simple
calculation shows that retaining only the linear order in the
amplitude of the distortion, the spin and the charge densities
exhibit spatial oscillations with wave vector $2k_F^+$. Specifically:

\begin{widetext}
\begin{eqnarray}
\frac{\delta S_x \left( \vec {r} \right)}{\hbar}\!\!\!&\!\! =\!\!&\!
\!\! \sum\limits_{\vec k}  A_{\vec k} \left[\left(\cos \phi _{\vec
{k} + \frac{\vec Q}{2}} - \cos \phi _{\vec k -\frac{\vec Q}{2}}
\right)\sin \vec {Q} \cdot \vec r +\left( {\sin \phi _{\vec {k} +
\frac{\vec {Q}}{2}} + \sin \phi _{\vec {k} - \frac{\vec {Q}}{2}} }
\right)\cos \vec {Q} \cdot \vec
{r}\right]~, \nonumber\\
\frac{\delta S_y \left( \vec r \right)}{\hbar}\!\! &\!\!=\!\!
&\!\!\! \sum\limits_k A_{\vec k} \left[\left( {\sin \phi _{\vec {k}
+ \frac{\vec {Q}}{2}} - \sin \phi _{\vec {k} - \frac{\vec {Q}}{2}} }
\right)\sin \vec {Q} \cdot \vec {r} -\left( \cos \phi _{\vec k +
\frac{\vec Q}{2}} + \cos \phi _{\vec k -
\frac{\vec Q}{2}}\right)\cos \vec Q \cdot \vec r  \right] ~,\nonumber\\
\frac{\delta S_z \left( \vec {r} \right)}{\hbar} \!\!&\!\!=\!\!&\!\!
\sum\limits_k 2 A_{\vec k } \left[ 1 - \cos \left( \phi _{\vec k +
\frac{\vec Q}{2}} - \phi _{\vec k - \frac{\vec Q}{2}}
\right) \right]\cos \vec {Q} \cdot \vec r  ~,\nonumber\\
\delta n \left( \vec {r} \right) \!\!&\!\!=\!\!&\!\! \sum\limits_k 4
A_{\vec k } \left[ 1 + \cos \left( \phi _{\vec k + \frac{\vec Q}{2}}
- \phi _{\vec k - \frac{\vec Q}{2}} \right) \right]\cos \vec {Q}
\cdot \vec r~.
\end{eqnarray}

The change in the HF energy is obtained by substituting the
expression of the non-zero density matrix elements from
Eq.~(\ref{delt_rho_limit}) into Eq.~(\ref{DeltaEHF}). The resulting
expression can be expressed as:

\begin{equation}
\label{DeltaEH} \Delta ^ {(2)}E_{HF}= \Delta _0^{(2)} E_{HF} +
\Delta _H^{(2)} E_{HF} + \Delta _X^{(2)} E_{HF}~,
\end{equation}
where we have defined

\begin{eqnarray}
\label{Delts_0} \Delta _0^{\left( 2 \right)} E_{HF} &=&
{\sum\limits_{\vec {k} } {\left( {n_{\vec {k} + \frac{\vec {Q}}{2}}
+ n_{\vec {k} - \frac{\vec {Q}}{2}} } \right)} \frac{\epsilon _{\vec
{k} - \frac{\vec {Q}}{2}, + } - \epsilon _{\vec {k} + \frac{\vec
{Q}}{2}, + } }{n_{\vec {k} + \frac{\vec
{Q}}{2}} - n_{\vec {k} - \frac{\vec {Q}}{2}} }A_k^2 }~,\\
\label{Delts_H}
\Delta _H^{\left( 2 \right)} E_{HF} &= &
{\frac{1}{2}\sum\limits_{\vec {k},\vec {p}} {\left( {v_{\vec {k} +
\frac{\vec {Q}}{2}, + ,\vec {p} - \frac{\vec {Q}}{2}, + ,\vec {p} +
\frac{\vec {Q}}{2}, + ,\vec {k} - \frac{\vec {Q}}{2}, + } + v_{\vec
{k} - \frac{\vec {Q}}{2}, + ,\vec {p} + \frac{\vec {Q}}{2}, + ,\vec
{p} - \frac{\vec {Q}}{2}, +
,\vec {k} + \frac{\vec {Q}}{2}, + } } \right)}} \nonumber\\
&\times &\left( {n_{\vec {k} + \frac{\vec {Q}}{2}} + n_{\vec {k} -
\frac{\vec {Q}}{2}} } \right)\left( {n_{\vec {p} + \frac{\vec
{Q}}{2}} + n_{\vec {p} - \frac{\vec
{Q}}{2}} } \right)A_{\vec {k}} A_{\vec {p}}~, \\
\label{Delts_X}
\Delta _X^{\left( 2 \right)E_{HF}} &=& - {\frac{1}{2}\sum\limits_{\vec
{k},\vec {p}} {\left( {v_{\vec {k} + \frac{\vec {Q}}{2}, + ,\vec {p}
- \frac{\vec {Q}}{2}, + ,\vec {k} - \frac{\vec {Q}}{2}, + ,\vec {p}
+ \frac{\vec {Q}}{2}, + } + v_{\vec {k} - \frac{\vec {Q}}{2}, +
,\vec {p} + \frac{\vec {Q}}{2}, + ,\vec {k} + \frac{\vec {Q}}{2}, +
,\vec {p} - \frac{\vec {Q}}{2}, + } } \right)}} \nonumber\\ &\times&
\left( {n_{\vec {k} + \frac{\vec {Q}}{2}} + n_{\vec {k} - \frac{\vec
{Q}}{2}} } \right)\left( {n_{\vec {p} + \frac{\vec {Q}}{2}} +
n_{\vec {p} - \frac{\vec {Q}}{2}} } \right)A_{\vec {k}} A_{\vec {p}
}~.
\end{eqnarray}

The Hartree and exchange terms in Eqs.~(\ref{Delts_H})-(\ref{Delts_X})
contain combinations of the matrix elements of the electron-electron
interaction. By employing Eq.~(\ref{phiSO}), after simple algebraic
manipulations, we obtain:

\begin{eqnarray}
\label{int_mat_el_sum_limitH} v_{\vec {k} + \frac{\vec {Q}}{2}, +
,\vec {p} - \frac{\vec {Q}}{2}, + ,\vec {p} + \frac{\vec {Q}}{2}, +
,\vec {k} - \frac{\vec {Q}}{2}, + } &+ &v_{\vec {k} - \frac{\vec
{Q}}{2}, + ,\vec {p} + \frac{\vec {Q}}{2}, + ,\vec {p} - \frac{\vec
{Q}}{2}, + ,\vec {k} + \frac{\vec {Q}}{2}, + } =
\nonumber\\
\frac{v_Q }{2L^2}\left[ 1+ \cos \left( \phi _{\vec {k} +\frac{\vec
{Q}}{2}} - \phi _{\vec {k} - \frac{\vec Q}{2}} \right)+ \cos
\left(\phi _{\vec p + \frac{\vec Q}{2}} - \phi _{\vec p - \frac{\vec
Q}{2}} \right)\right.& +&\left. \cos \left( \phi _{\vec k +
\frac{\vec Q}{2}} - \phi _{\vec k - \frac{\vec Q}{2}} + \phi _{\vec
p + \frac{\vec Q}{2}} - \phi _{\vec p - \frac{\vec Q}{2}} \right)
\right]~,
\end{eqnarray}
and
\begin{eqnarray}
\label{int_mat_el_sum_limitE} v_{\vec {k} + \frac{\vec {Q}}{2}, +
,\vec {p} - \frac{\vec {Q}}{2}, + ,\vec {k} - \frac{\vec {Q}}{2}, +
\vec {p} + \frac{\vec {Q}}{2}, + } &+& v_{\vec {k} - \frac{\vec
{Q}}{2}, + ,\vec {p} + \frac{\vec {Q}}{2}, + ,\vec {k} +
\frac{\vec {Q}}{2}, + ,\vec {p} - \frac{\vec {Q}}{2}, + } = \nonumber\\
\frac{v_{\left| {\vec {k} - \vec {p}} \right|} }{2L^2}\left[ 1 +
\cos \left( \phi _{\vec p + \frac{\vec Q}{2}} - \phi _{\vec k +
\frac{\vec Q}{2}}\right) + \cos \left( \phi _{\vec k - \frac{\vec
Q}{2}} - \phi _{\vec p - \frac{\vec Q}{2}} \right) \right.&+&\left.
\cos \left(\phi _{\vec p + \frac{\vec Q}{2}} - \phi _{\vec {k} +
\frac{\vec Q}{2}} + \phi _{\vec k - \frac{\vec Q}{2}} - \phi _{\vec
p - \frac{\vec Q}{2}} \right) \right]~.
\end{eqnarray}

In order to explicitly evaluate the change in the Hartree-Fock
energy, we need to assume a specific expression for the distortion
amplitudes. As a first condition, we will perturb only a narrow
region near the Fermi surface. Following the same pattern of the proof
of reference \cite{TheBook}, we propose for the present problem the following
educated variational guess:

\begin{equation}
\label{amplitude} A_{\vec {k}} = \left\{
\begin{array} {ll}
\frac{\left( bk_F^+ \right)^{\frac{3}{2}}} {\ln \frac{2}{b}}
\frac{\left| n_{\vec k + \frac{\vec Q}{2}} - n_{\vec k - \frac{\vec
Q}{2}} \right|} {k^{\frac{3}{2}}} \frac{\sqrt {\left| \sin \phi
_{\vec k}  \right| }} {\left| \cos \phi _{\vec k}  \right|}
,& bk_F^+< k < \varsigma bk_F^+  \\
 0, & $otherwise$~,
\end{array}  \right.
\end{equation}
where $b \ll 1$ is our small parameter and the second arbitrary $\varsigma > 1$. This
expression is intentionally chosen to display singularities for
$k=0$ and $\phi_k=0$. These singularities are crucial to the present
proof.

We now notice that in the expressions of the interaction matrix
elements, there appear factors of the type:

\begin{equation}
\label{cosphasesmall} \cos \left( {\phi _{\vec {p} + \frac{\vec
{Q}}{2}} - \phi _{\vec {k} + \frac{\vec {Q}}{2}} } \right) =
\frac{\left( {p\cos \phi _{\vec {p}} + k_F^+ } \right)\left( {k\cos
\phi _{\vec {k} }+ k_F^+ } \right) + pk\sin \phi _{\vec {p}} \sin
\phi _{\vec {k} }}{\sqrt {p^2 + (k_F^+)^2 + 2k_F^+ p\cos \phi _{\vec
{p}} } \sqrt {k^2 + (k_F^+)^2 + 2k_F^+ k\cos \phi _{\vec {k} }} }~.
\end{equation}
\end{widetext}
Since we are only interested in the leading order expansion with
respect to $b$, these cosines can be simply taken to be equal to
unity, since in the region where the amplitude of the distortion is
non-vanishing, both $p/k_F^+$ and $k/k_F^+$ are of order $b$.
Accordingly we will assume
\begin{equation}
\label{cosphasesdiff} \cos \left( {\phi _{\vec {p} + \frac{\vec
{Q}}{2}} - \phi _{\vec {k} + \frac{\vec {Q}}{2}} } \right) \simeq 1
+ O\left( {b ^2} \right)~.
\end{equation}

At this point, we recall that $\vec k + \frac{\vec Q}{2}$ and $\vec
k - \frac{\vec Q}{2}$ must lie on opposite sides of the Fermi sea
(i.e $|\vec k - \frac{\vec Q}{2}| < k_F^+$ and $|\vec k + \frac{\vec
Q}{2}| > k_F^+$ ), which implies that $\left| \cos \phi_{\vec k}
\right|> \frac{b}{2}$.

We can now proceed to the evaluation of the three components of
$\Delta ^{\left( 2 \right)}E_{HF} $ from Eqs.~(\ref{Delts_H})-(\ref{Delts_X})
to leading order in $b$.

For $\Delta _0^{\left( 2 \right)} E_{HF} $ the first step is to
calculate $\epsilon _{\vec {k} - \frac{\vec {Q}}{2}, + } - \epsilon
_{\vec {k} + \frac{\vec {Q}}{2}, + } $. Using (\ref{EHF}), with
energies expressed in Ry, $x = \frac{k}{k_F }$ and $x' =
\frac{k'}{k_F }$, we have:

\begin{widetext}
\begin{eqnarray}
\label{epsilon00} \epsilon _{\vec {k} - \frac{\vec {Q}}{2}} -
\epsilon _{\vec {k} + \frac{\vec {Q}}{2}} &=& \left. {\left(
{\frac{4u^2}{r_s^2 } - \frac{4\tilde {\alpha }u}{r_s } -
\frac{1}{\pi r_s }\int\limits_0^{2\pi } {\int\limits_0^1
{\frac{{x}'\left( {1 + \cos {\varphi }'} \right)d{x}'d{\phi
}'}{\sqrt {u^2 + {x}'^2 - 2u{x}'\cos {\varphi }'} }} } } \right)}
\right|_{u = \sqrt
{1 + x^2 - 2x\cos \phi_{\vec k} } } \nonumber\\
 &-& \left. {\left( {\frac{4u^2}{r_s^2 } - \frac{4\tilde {\alpha }u}{r_s } -
\frac{1}{\pi r_s }\int\limits_0^{2\pi } {\int\limits_0^1
{\frac{{x}'\left( {1 + \cos {\varphi }'} \right)d{x}'d{\phi
}'}{\sqrt {u^2 + {x}'^2 - 2u{x}'\cos {\varphi }'} }} } } \right)}
\right|_{u =
\sqrt {1 + x^2 + 2x\cos \phi_{\vec k} } } ~. \nonumber\\
 \end{eqnarray}
\end{widetext}
The quadratures appearing in this expression can then be manipulated
by making use of the results of Appendix \ref{Elliptic}. The result is:
\begin{equation}
\label{epsilon0} \epsilon _{\vec {k} - \frac{\vec {Q}}{2}} -
\epsilon _{\vec {k} + \frac{\vec {Q}}{2}} = - \frac{16x\cos \phi
}{r_s^2 }\left( {1 - \frac{\tilde {\alpha }r_s }{2} - \frac{r_s
}{2\pi }\ln \left| {\xi x\cos \phi } \right|} \right)~,
\end{equation}
where the logarithmic term accounts for the divergence of the
derivative of the HF single particle energy near the Fermi level.
Here, $\xi$ is a constant approximately equal to $0.51$.

By substituting (\ref{epsilon0}) and (\ref{amplitude}) in (\ref{Delts_0})
we obtain:
\begin{widetext}
\begin{equation}
\Delta _0^{\left( 2 \right)} E_{HF} = \frac{16b^3L^2k_F^2 }{\left(
{\ln \frac{2}{b}} \right)^2r_s^2 \pi ^2}\int\limits_b^{\varsigma b}
{\frac{dx}{x}\int\limits_0^{\arccos \frac{b}{2}} {d\varphi
\frac{\sin \varphi }{\cos \varphi }\left( {1 - \frac{\tilde {\alpha
}r_s }{2} - \frac{r_s }{2\pi }\ln \left| {\xi x\cos \varphi }
\right|} \right)} }~,
\end{equation}
\end{widetext}
an expression that, to leading order in $b$, reduces to:
\begin{equation}
\Delta _0^{\left( 2 \right)} E_{HF} \simeq \frac{48Nb^3\ln \varsigma
}{r_s \pi ^2}~.
\end{equation}
where $N$ is the number of particles.

The Hartree term $\Delta _H^{\left( 2 \right)} E_{HF} $ (containing
$v_Q$) can in turn be evaluated as follows. By making use of the assumed
amplitude in (\ref{amplitude}), we write:
\begin{widetext}
\begin{equation}
\Delta _H^{(2)} E_{HF} = \frac{16Nb^3}{\pi ^2\left( {\ln
\frac{2}{b}} \right)^2r_s }\int\limits_b^{b\varsigma }
{\frac{dx}{\sqrt x }\int\limits_b^{b\varsigma } {\frac{dy}{\sqrt y
}\int\limits_0^{\arccos \frac{b}{2}} {d\varphi _x \frac{\sqrt {\sin
\varphi _x } }{\cos \phi _x }\int\limits_0^{\arccos \frac{b}{2}}
{d\varphi _y \frac{\sqrt {\sin \varphi _y } }{\cos \varphi _y }} } }
}~.
\end{equation}
\end{widetext}
At this point, using the result (\ref{Rq1}), the leading order in
$b$ of this quantity is given by:
\begin{equation}
\Delta _H^{\left( 2 \right)} E_{HF} \simeq \frac{64Nb^4\left( {\sqrt
\varsigma - 1} \right)^2}{\pi ^2r_s }~.
\end{equation}

The last term of (\ref{DeltaEH}), the exchange energy contribution,
is clearly negative and therefore will certainly lower the energy.
Its evaluation is formidable, for it involves several complicated
and seemingly difficult quadratures. Rather than attempting to actually
calculate it, we will establish a lower limit for its magnitude.

We will restrict ourselves to the region in which both angles $\phi
_{\vec {k}} $ and $\phi _{\vec {p}} $ are in the first quadrant.
Since this excludes some contributions of the same sign,
the exchange energy will be underestimated. We therefore have:
\begin{widetext}
\begin{equation}
\label{EXHF} \left|\Delta _X^{\left( 2 \right)} E_{HF}\right| \geq
\frac{2Nb^3}{\pi ^2\left( {\ln \frac{2}{b}} \right)^2r_s
}\int\limits_b^{b\varsigma } {\!dx\int\limits_b^{b\varsigma }
{\!dy\!\!\int\limits_0^{\arccos \frac{b}{2}} {\!\!d\varphi _x
\int\limits_0^{\arccos \frac{b}{2}} {\!\!d\varphi _y \frac{1}{\sqrt
{x^2 + y^2 - 2xy\cos \left( \varphi _{x} - \varphi _{y } \right)
}}}}}}
 \frac{1}{\sqrt x
}\frac{\sqrt {\left| {\sin \varphi _x } \right|} }{\left| {\cos
\varphi _x } \right|}\frac{1}{\sqrt y }\frac{\sqrt {\left| {\sin
\varphi _y } \right|} }{\left| {\cos \varphi _y } \right|}~.
\end{equation}
\end{widetext}

It is simple to see that this expression will turn out to be
proportional to $\left( {\ln b} \right)^3$. This is due to the
presence of three singularities in the denominator of the integrand.
Of these one stems from the divergence of the Coulomb potential,
while the other two come from the upper limit of the angular
integrations.

Another simplification is provided by the use of the inequality:
\begin{equation}
\frac{1}{\sqrt {x^2 + y^2 - 2xy\cos \left( \varphi _x - \varphi _y
\right)} } \ge \frac{1}{\sqrt {x^2 + y^2 - 2xy\sin \varphi _x\sin
\varphi_y } } ~.
\end{equation}

Using the same changes of variable as in Appendix \ref{Special},
i.e. $t_x=\tan(\phi_x/2)$ and $t_y=\tan(\phi_y/2)$, we can rewrite
this integral as follows:
\begin{widetext}
\begin{equation}
\left|\Delta _X^{\left( 2 \right)} E_{HF}\right| \geq
\frac{16Nb^3}{\pi ^2\left( {\ln \frac{2}{b}} \right)^2r_s }
\int\limits_b^{b\varsigma } {\frac{dx}{\sqrt x }
\int\limits_b^{b\varsigma } {\frac{dy}{\sqrt y }
\int\limits_0^{\sqrt{ \frac{2-b}{2+b}}}\! dt_x\!
\int\limits_0^{\sqrt{ \frac{2-b}{2+b}}} dt_y \frac{1}{\sqrt {x^2 +
y^2 -  \frac{8xyt_x t_y}{(1+t_x^2)(1+t_y^2)}} } } }
\frac{1}{1-t_x^2} \sqrt{\frac{t_x}{1+t_x^2}} \frac{1}{1-t_y^2}
\sqrt{\frac{t_y}{1+t_y^2}}.
\end{equation}
\end{widetext}
It is clear now that the main contribution to the integral comes
from the region around the upper limit of integration for both $t_x$
and $t_y$. In order to retain the leading order term, a good
approximation will be to replace both $t$'s with $1 - b/2$ in the
denominator of the first square root. In this way, we can separate
the angular integrations in (\ref{EXHF}) to obtain:
\begin{widetext}
\begin{equation}
\left|\Delta _X^{\left( 2 \right)} E_{HF}\right|\geq\frac{2Nb^3}{\pi
^2\left( {\ln \frac{2}{b}} \right)^2r_s }\int\limits_b^{b\varsigma }
{\frac{dx}{\sqrt x }\int\limits_b^{b\varsigma } {\frac{dy}{\sqrt y
}\frac{1}{\sqrt {x^2 + y^2 - 2xy\left( {1 -
\frac{b^2}{8}} \right)^2} } } }
\int\limits_0^{1-\frac{b}{2}} dt_x \int\limits_0^{1-\frac{b}{2}}
dt_y \frac{1}{1-t_x^2} \sqrt{\frac{t_x}{1+t_x^2}} \frac{1}{1-t_y^2}
\sqrt{\frac{t_y}{1+t_y^2}}~.
\end{equation}
\end{widetext}

The last two integrals are evaluated using (\ref{Rq1}) and other
changes of variables ($x=ub$, $y=vb$)  :
\begin{equation}
\left|\Delta _X^{\left( 2 \right)} E_{HF}\right|\geq
\frac{8Nb^3}{\pi ^2r_s }\int\limits_1^{\sqrt \varsigma }
{du\int\limits_1^{\sqrt \varsigma } {dv\frac{1}{\sqrt {\left( {u^2 -
v^2} \right)^2 + \frac{u^2v^2b^2}{2}} }} }~.
\end{equation}

The last quadrature is calculated in (\ref{Rq2}), leading us to a
very simple inequality for the exchange contribution:
\begin{equation}
\left|\Delta _X^{\left( 2 \right)} E_{HF}\right|\geq\frac{8Nb^3}{\pi
^2r_s }\ln \varsigma \ln \frac{1}{b}~.
\end{equation}
This term contains a logarithmic factor $\ln \frac{1}{b}$, which
allows the negative change in the exchange contribution to control
all the remaining terms. This concludes the proof for this case.

\begin{figure}[h]
\centerline{\includegraphics[width=3.0in,height=2.0in]{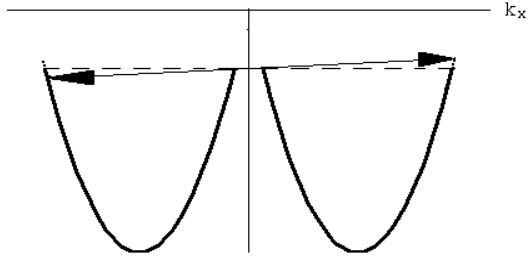}}
\caption{  \label{Highrscouplig} HF instability: symmetry breaking
coupling for the case in which only the lower subband is occupied.}
\end{figure}

The same chain of arguments does apply to the case in which the
generalized chirality is greater than one. The coupling that produces
this kind of instability is schematically shown in Fig.~\ref{Highrscouplig}.
All the formulas we derived in the previous case do still apply. The
only difference lies in the lower integration limits of
Eq.~(\ref{epsilon00}), but no relevant contribution ensues from this.
The matrix elements related to the Hartree and exchange contributions
are the same, and, as a consequence, the leading order approximation
is identical.

\subsection{Instability for the case of chirality less than one}
\label{chiless1}
The argument of the previous Section can be applied when the
chirality is less than one, i.e. when both chiral subbands are
occupied. We can try to break the symmetry by coupling states with
the same chirality as well as states with different chiralities.
When same chirality states are coupled, there is nothing new, as one
simply just adds a chirality index to the various quantities. In
this case, the wave vectors characterizing the oscillations are given
by: $Q_{\mu}=2k_F^{\mu}$ and the trial wavefuntions can be written
as:

\begin{equation}
\Psi _{\vec {k}\mu } \simeq \psi _{\vec {k}\mu } + A_{\vec
{k} + \vec {Q}_\mu ,\vec {k}\mu } \psi _{\vec {k} + \vec
{Q}_\mu \mu } + A_{\vec {k} - \vec {Q}_\mu ,\vec {k}\mu }
\psi _{\vec {k} - \vec {Q}_\mu \mu }~.
\end{equation}

\begin{figure}[h]
\centerline{\includegraphics[width=3.0in,height=2.0in]{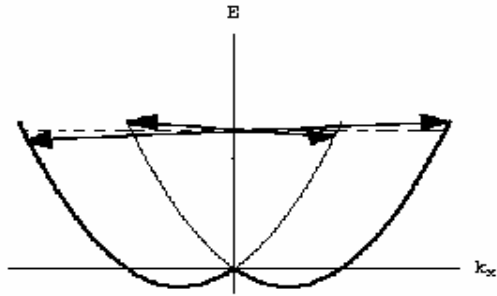}}
\caption{ \label{Lowrscoupling} Symmetry breaking coupling of states
with the same chirality.}
\end{figure}

This type of coupling is depicted in Fig. \ref{Lowrscoupling}.

The corresponding distortion of the components of the spin density
and the number density can be again calculated up to the first order
in the amplitudes:

\begin{widetext}
\begin{eqnarray}
\delta S_x \left( \vec {r} \right) &=& \hbar \left[
{\sum\limits_{k,\mu } {A_{\vec {k}\mu } \left( {\cos \phi _{\vec {k}
+ \frac{\vec {Q}_\mu }{2}} - \cos \phi _{\vec {k} - \frac{\vec
{Q}_\mu }{2}} } \right)\sin \vec
{Q}_\mu \cdot \vec {r}} }
+\sum\limits_{k,\mu } {A_{\vec {k}\mu } \left( {\sin \phi _{\vec {k}
+ \frac{\vec {Q}_\mu }{2}} + \sin \phi _{\vec {k} - \frac{\vec
{Q}_\mu }{2}} } \right)\cos \vec {Q}_\mu \cdot \vec {r}}
\right]~,\\
\delta S_y \left( \vec {r} \right)& =& \hbar \left[
\sum\limits_{k,\mu } {A_{\vec {k}\mu } \left( {\sin \phi _{\vec {k}
+ \frac{\vec {Q}_\mu }{2}} - \sin \phi _{\vec {k} - \frac{\vec
{Q}_\mu }{2}} } \right)\sin \vec {Q}_\mu \cdot \vec {r}} -
\sum\limits_{k,\mu } {A_{\vec {k}\mu } \left( {\cos \phi _{\vec {k}
+ \frac{\vec {Q}_\mu }{2}} + \cos \phi _{\vec {k} - \frac{\vec
{Q}_\mu }{2}} } \right)\cos \vec {Q}_\mu \cdot \vec {r}}
\right]~,\\
\delta S_z \left( \vec {r} \right) &=& 2\hbar \left[
{\sum\limits_{k,\mu }  {A_{\vec {k}\mu } \left( {1 - \cos \left(
{\phi _{\vec {k} + \frac{\vec {Q}_\mu }{2}} - \phi _{\vec {k} -
\frac{\vec {Q}_\mu }{2}} } \right)} \right)\cos \vec {Q}_\mu \cdot
\vec {r}} } \right]~, \\
\delta n \left( \vec {r} \right) &=& 4  {\sum\limits_{k,\mu }
{A_{\vec {k}\mu } \left( {1 - \cos \left( {\phi _{\vec {k} +
\frac{\vec {Q}_\mu }{2}} - \phi _{\vec {k} - \frac{\vec {Q}_\mu
}{2}} } \right)} \right)\cos \vec {Q}_\mu \cdot \vec {r}} }~.
\end{eqnarray}
\end{widetext}
We proceed in this case by choosing an amplitude not unlike the one
assumed above:
\begin{equation}
A_{\vec {k}\mu } = \left\{
\begin{array}{ll}
\frac{\left(bk_F^{\mu}  \right)^{\frac{3}{2}}}{\ln \frac{2}{b}}
\frac{\left| n_{\vec {k} + \frac{\vec {Q}}{2}\mu } - n_{\vec {k} -
\frac{\vec {Q}}{2}\mu }\right|} {k^{\frac{3}{2}}} \frac{\sqrt
{\left| {\sin \phi _{\vec {k}} } \right|} }{\left| {\cos \phi _{\vec
{k} }}\right|}
,& bk_F^{\mu} < k < \varsigma bk_F^\mu  \\
0,& $otherwise$~.\\
\end{array} \right.
\end{equation}
The proof of the corresponding instability theorem proceeds then in
exactly the same manner.

\begin{figure}[h]
\centerline{\includegraphics[width=3.0in,height=2.0in]{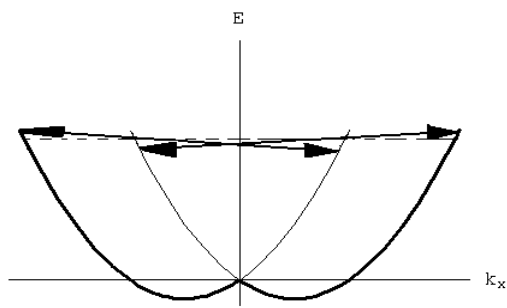}}
\caption{  \label{Diffchir} States with opposite chiralities coupled
to obtain a HF instability}
\end{figure}

As anticipated, there is also not much difference when we try to
couple states with opposite chirality (see Fig. \ref{Diffchir}).
Although some of the expressions involved in the derivation do
change, the main features of the argument remain unchanged.
The coupling vector in this case is given by $Q=k_F^{+}+k_F^{-}$.
Here, we try to find a lower energy state by coupling wavefunctions
with wave vector $\vec {k}$ with those with wave vector $\vec {k}\pm
\vec {Q}$ and opposite chirality. The trial wavefunctions then read:
\begin{equation}
\label{amplitudeopp} \Psi _{\vec {k} + } \simeq \psi _{\vec
{k} + } + A_{\vec {k} + \vec {Q},\vec {k}} \psi
_{\vec {k} + \vec {Q}, - } + A_{\vec {k} - \vec {Q},\vec
{k}\mu } \psi _{\vec {k} - \vec {Q}, - }~.
\end{equation}

For $A_{\vec {k}\pm \vec {Q},\vec {k}} =
A_{\vec {k},\vec {k}\pm \vec {Q}} $ the only non zero variations
of the matrix elements of the single-particle density matrix
operator acquire the following form:

\begin{equation}
\label{delta_rho_opposite} \delta \rho _{\vec {k} + \frac{\vec
{Q}}{2},\mu ,\vec {k} - \frac{\vec {Q}}{2}, - \mu } = \delta \rho
_{\vec {k} - \frac{\vec {Q}}{2},\mu ,\vec {k} + \frac{\vec {Q}}{2},
- \mu } = \left( {n_{\vec {k} + \frac{\vec {Q}}{2},\mu } + n_{\vec
{k} - \frac{\vec {Q}}{2}, - \mu } } \right)A_{\vec {k}}~.
\end{equation}

\begin{widetext}
In this case, the new state is characterized by a similar spin and
density modulation:

\begin{eqnarray}
\delta S_x \left( \vec r \right) &=& \hbar \sum\limits _k A_{\vec k}
\left[ \cos \left( \vec Q \cdot \vec r - \phi _{\vec k + \frac{\vec
Q}{2}} - \phi _{\vec k - \frac{\vec Q}{2}}
\right) + \cos \vec Q \cdot \vec r \right]~, \\
\delta S_y \left( \vec r \right) &=& \hbar \sum\limits _k A_{\vec k}
\left[ \sin \left( -\vec Q \cdot \vec r + \phi _{\vec k + \frac{\vec
Q}{2}} + \phi _{\vec k - \frac{\vec Q}{2}}
\right) + \sin \vec Q \cdot \vec r \right]~,\\
\delta S_z \left( \vec r \right) &=& \hbar \sum\limits _k A_{\vec k}
\left[ \sin \left( \vec Q \cdot \vec r - \phi _{\vec k - \frac{\vec
Q}{2}} \right) + \sin \left(\vec Q \cdot \vec r - \phi
_{\vec k + \frac{\vec Q}{2}} \right)\right]~, \\
\delta n \left( \vec r \right) &=& 2 \sum\limits _k A_{\vec k}
\left[ \sin \left( \vec Q \cdot \vec r - \phi _{\vec k - \frac{\vec
Q}{2}} \right) - \sin \left(\vec Q \cdot \vec r - \phi _{\vec k +
\frac{\vec Q}{2}} \right)\right]~.
\end{eqnarray}

The corresponding terms in the Hartree-Fock energy change are:

\begin{eqnarray}
\Delta_0 ^{\left( 2 \right)}E_{HF} &=&{\sum\limits_{\vec {k}\mu }
{\left( {n_{\vec {k} + \frac{\vec {Q}}{2},\mu } + n_{\vec {k} -
\frac{\vec {Q}}{2}, - \mu } } \right)\frac{\epsilon _{\vec {k} -
\frac{\vec {Q}}{2}\mu } - \epsilon _{\vec {k} + \frac{\vec {Q}}{2} -
\mu } }{n_{\vec {k} + \frac{\vec {Q}}{2} - \mu } - n_{\vec {k} -
\frac{\vec {Q}}{2}\mu }
}A_k^2 } }~, \\
\Delta_H ^{\left( 2 \right)}E_{HF}& =&\frac{1}{2}\sum\limits_{\vec
{k}\vec {p}\mu \nu} {\left( {v_{\vec {k} + \frac{\vec {Q}}{2},\mu
,\vec {p} - \frac{\vec {Q}}{2}\nu ,\vec {p} + \frac{\vec {Q}}{2}, -
\nu ,\vec {k} - \frac{\vec {Q}}{2}, - \mu } + v_{\vec {k} -
\frac{\vec {Q}}{2},\mu ,\vec {p} + \frac{\vec {Q}}{2},\nu ,\vec {p}
- \frac{\vec {Q}}{2}, - \nu ,\vec {k} + \frac{\vec {Q}}{2}, - \mu }
} \right)} \nonumber\\
&&\left( {n_{\vec {k} + \frac{\vec {Q}}{2},\mu } + n_{\vec {k} -
\frac{\vec {Q}}{2}, - \mu } } \right)\left( {n_{\vec {p} +
\frac{\vec {Q}}{2}\nu } + n_{\vec {p} - \frac{\vec {Q}}{2} -
\nu } } \right)A_{\vec {k}\mu } A_{\vec {p}\nu }~,\\
\Delta _X^{\left( 2 \right)} E_{HF}&=& - \frac{1}{2}\sum\limits_{\vec
{k}\vec {p} \mu \nu} {\left( {v_{\vec {k} + \frac{\vec {Q}}{2}\mu
,\vec {p} - \frac{\vec {Q}}{2}, - \nu ,\vec {k} - \frac{\vec
{Q}}{2}, - \mu ,\vec {p} + \frac{\vec {Q}}{2},\nu } + v_{\vec {k} -
\frac{\vec {Q}}{2}, - \mu ,\vec {p} + \frac{\vec {Q}}{2},\nu ,\vec
{k} + \frac{\vec {Q}}{2}\mu ,\vec {p} - \frac{\vec {Q}}{2}, - \nu }
} \right)} \nonumber \\
&&\left( {n_{\vec {k} + \frac{\vec {Q}}{2},\mu } + n_{\vec {k} -
\frac{\vec {Q}}{2}, - \mu } } \right)\left( {n_{\vec {p} +
\frac{\vec {Q}}{2},\nu } + n_{\vec {p} - \frac{\vec {Q}}{2}, - \nu }
} \right)A_{\vec {k}\mu } A_{\vec {p}\nu }~.
\end{eqnarray}
\end{widetext}
The relevant interaction matrix elements which appear in the
expression of the Hartree term become:
\begin{eqnarray}
\frac{2v_Q }{L^2}\left( {\sin \phi _{\vec {p} + \frac{\vec {Q}}{2}}
- \sin \phi _{\vec {p} - \frac{\vec {Q}}{2}} } \right)\left( {\sin
\phi _{\vec {k} - \frac{\vec {Q}}{2}} - \phi _{\vec {k} + \frac{\vec
{Q}}{2}} } \right) ~,
\end{eqnarray}
while those determining the exchange contribution reduce to:

\begin{eqnarray}
\frac{v_{|\vec k - \vec p|} }{L^2}\left( \cos\left( { \phi _{\vec
{p} + \frac{\vec {Q}}{2}} - \phi _{\vec {k} + \frac{\vec {Q}}{2}} }
\right)+ \cos\left( { \phi _{\vec {p} - \frac{\vec {Q}}{2}} - \phi
_{\vec {k} - \frac{\vec {Q}}{2}} } \right)  + \right.\nonumber\\
+  \cos\left( { \phi _{\vec {p} + \frac{\vec {Q}}{2}} - \phi _{\vec
{k} - \frac{\vec {Q}}{2}} } \right)+ \cos\left( { \phi _{\vec {p} -
\frac{\vec {Q}}{2}} - \phi _{\vec {k} + \frac{\vec {Q}}{2}} }
\right)- \nonumber\\
-\cos\left( { \phi _{\vec {k} + \frac{\vec {Q}}{2}} - \phi _{\vec
{k} - \frac{\vec {Q}}{2}} } \right) - \cos \left( {\phi _{\vec {p} +
\frac{\vec {Q}}{2}} - \phi
_{\vec {p} - \frac{\vec {Q}}{2}} } \right) +\nonumber\\
\left.+ \cos\left( { \phi _{\vec {p} + \frac{\vec {Q}}{2}} - \phi
_{\vec {k} + \frac{\vec {Q}}{2}} + \phi _{\vec {k} - \frac{\vec
{Q}}{2}} - \phi _{\vec {p} - \frac{\vec {Q}}{2}}} \right)+1
\right)~.
\end{eqnarray}

As in the previous cases, we assume the following coupling
amplitude:

\begin{equation}
A_{\vec {k}}^{\mu} = \left\{
\begin{array}{ll}
\frac{\left( bk_F^ {\mu}  \right)^{\frac{3}{2}}} {\ln\frac{2}{b}}
\frac{\left| n_{\vec k + \frac{\vec {Q}}{2}, \mu } - n_{\vec k -
\frac{\vec {Q}}{2}, -\mu }  \right| }{k^{\frac{3}{2}}} \frac{\sqrt
{\left| \sin \phi _{\vec {k} } \right|} }{\left| \cos \phi _{\vec k}
\right|},& bk_F^ + < k <
\varsigma bk_F^ + ,|\cos \phi_{\vec k}|<\frac{b}{2}  \\
0,& $otherwise$~. \\
\end{array} \right.
\end{equation}

The first term in the Hartree-Fock energy (\ref{Delts_0})
has a positive contribution which is proportional to $b^3$. This
originates from the same logarithmic factor associated with the
divergence of the derivative of the single-particle energy at the
Fermi level. The Hartree term introduces higher order terms in $b$
due to the presence of the sine factors in its matrix elements.
Finally, the leading order contribution to the exchange matrix
elements is $4v_{\left| {\vec {k} - \vec {p}} \right|} $. By evaluating
integrals similar to those in Eq.~(\ref{EXHF}) we again obtain a
negative energy change of order $b^3\ln (1/b)$.

\section{Conclusions}
\label{conclusion}
We have been able to formally construct a number of distorted chiral states which, irrespective
of the electron density, have, within mean field, a lower energy than the spatially
homogeneous paramagnetic chiral HF ground state, thereby affording a rigorous
proof of a generalization of Overhauser's Hartree-Fock instability theorem to a
two dimensional electron liquid in the presence of linear Rashba-Dresselhaus spin-orbit coupling.
It is important to notice that, as mentioned in Section~\ref{RashbaHF}, to establish
the instability we have not allowed for momentum space repopulation: inclusion of this
phenomenon would have further lowered the energy of the trial states while greatly increasing
the difficulty of the analysis.

It is worth mentioning that the states that have been analyzed in
this paper differ in a number of ways form the original spin/charge
density waves proposed by Overhauser. Our states are chiral density
states, that display both spin and charge modulations. The presence
of charge modulations cannot be ignored (as commonly done in the case without
spin-orbit) and the Hartree term has to be evaluated explicitly.
The exchange gain is shown to be larger than the kinetic energy plus the ensuing Hartree terms.
In this respect the difference with the plain electron liquid is quite marked for
it is precisely the electrostatic term that is believed to be fatal to the
charge density waves in that case. We prove that this is not the case for the chiral waves.
The main reason is that the exchange energy behaves quite differently
as compared to that in the absence of spin-orbit interaction. In particular it
does not necessarily favor a homogeneous chiral instability as shown in Ref.
\cite{ChesiGFG}. Our calculations show that in a vast class of inhomogeneous states,
the exchange energy can induce a chiral instability for all densities of the electron liquid.
Of particular importance is that the chiral states coupled by the type of mean fields
explored here do always have opposite spins. The fact that such couplings are possible
for all values of the spin-orbit coupling is another consequence of the non analyticity
of the properties of the electron liquid on this parameter.

Our results only represent the first step in understanding non
homogeneous states in this interesting many-body system. How the
inclusion of correlations will modify the HF scenario is of course a
most important question. Advances in this respect can in principle
be pursued by following the program outlined in Reference~\onlinecite{Over3}.
Such a study will require a vastly more complicated analysis.
An alternative route is to make use of perturbative techniques to establish the role of
correlation effects in the high density limit. Part of this program has been carried out in
Reference~\onlinecite{Chesi_GFG_RPA_to_be_published} for homogeneous states.

\acknowledgements
The authors would like to acknowledge useful
discussions with A. W. Overhauser. GS work was partially supported
by a grant from the Purdue Research Foundation.

\appendix
\section{Elliptic integral expansion}
\label{Elliptic}
For the purpose of our calculations the following expression must be
evaluated in the limit of small $u$:

\begin{eqnarray}
\left. {\int\limits_0^{2\pi } {\int\limits_0^1 {\frac{{x}'\left( {1
+ \cos \varphi } \right)d{x}'d\varphi }{\sqrt {z^2 + {x}'^2 -
2z{x}'\cos \varphi } }} } } \right|_{z = 1 + u} \nonumber\\
- \left.
{\int\limits_0^{2\pi } {\int\limits_0^1 {\frac{{x}'\left( {1 + \cos
\varphi } \right)d{x}'d\varphi }{\sqrt {z^2 + {x}'^2 - 2z{x}'\cos
\varphi } }} } } \right|_{z = 1 - u} .
\end{eqnarray}

An obvious problem is the presence of singularities in the integrand
for $z = 1$.

Let us define:

\begin{equation}
A\left( z \right) = \int\limits_0^{2\pi } {\int\limits_0^1
{\frac{{x}'\left( {1 + \cos \varphi } \right)d{x}'d\varphi }{\sqrt
{z^2 + {x}'^2 - 2z{x}'\cos \varphi } }} }~.
\end{equation}

Of course, $A$ is not differentiable in $z = 1$. Still we have to
evaluate $\frac{\partial A\left( z \right)}{\partial z}$ and expand
it in an asymptotic series around $z = 1$.

\begin{eqnarray}
\frac{\partial }{\partial z}A\left( z \right)& =& \frac{\partial
}{\partial z}\left( {z\int\limits_0^{2\pi } {\int\limits_0^{1
\mathord{\left/ {\vphantom {1 z}} \right. \kern-\nulldelimiterspace}
z} {\frac{y\left( {1 + \cos \varphi } \right)dyd\varphi }{\sqrt {1 +
y^2 - 2y\cos \varphi } }} } }
\right)\nonumber \\
& = &\int\limits_0^{2\pi } {\int\limits_0^{1 \mathord{\left/
{\vphantom {1 z}} \right. \kern-\nulldelimiterspace} z}
{\frac{y\left( {1 + \cos \varphi } \right)dyd\varphi }{\sqrt {1 +
y^2 - 2y\cos \varphi } }} } \nonumber\\
&+& z\frac{\partial }{\partial z}\left( {\int\limits_0^{2\pi }
{\int\limits_0^{1 \mathord{\left/ {\vphantom {1 z}} \right.
\kern-\nulldelimiterspace} z} {\frac{y\left( {1 + \cos \varphi }
\right)dyd\varphi }{\sqrt {1 + y^2 - 2y\cos \varphi } }} } }
\right)~.
\end{eqnarray}

The derivative in the second term can be evaluated in the following
way:

\begin{eqnarray}
&&\frac{\partial }{\partial z}\left( {\int\limits_0^{2\pi }
{\int\limits_0^{1 \mathord{\left/ {\vphantom {1 z}} \right.
\kern-\nulldelimiterspace} z} {\frac{y\left( {1 + \cos \varphi }
\right)dyd\varphi }{\sqrt {1 + y^2 - 2y\cos \varphi } }} } }
\right)\nonumber\\
&=&- \frac{1}{z^2}\left. {\int\limits_0^{2\pi } {\frac{y\left( {1 +
\cos \varphi } \right)d\varphi }{\sqrt {1 + y^2 -
2y\cos \varphi } }} } \right|_{y = \frac{1}{z}} =\nonumber\\
&=&-\frac{1}{z^2}\int\limits_0^{2\pi } {\frac{\left( {1 + \cos
\varphi } \right)d\varphi }{\sqrt {1 + z^2 - 2z\cos \varphi } }}~.
\end{eqnarray}

We can integrate this expression over its angular part so that the
result is expressed as:

\begin{eqnarray}
\int\limits_0^{2\pi } {\frac{\left( {1 + \cos \varphi }
\right)d\varphi }{\sqrt {1 + z^2 - 2z\cos \varphi } }}& =& -
\frac{2\left| {z - 1} \right|}{z}E\left( {\frac{ - 4z}{\left( {z -
1} \right)^2}} \right) \nonumber\\
&+& \frac{2}{z}\frac{\left( {z + 1} \right)^2}{\left| {z - 1}
\right|}K\left( {\frac{ - 4z}{\left( {z - 1} \right)^2}}
\right)~,\nonumber\\
\end{eqnarray}
where $E$ and $K$ are the elliptic integrals of first and second
kind defined as in \cite{Wolfram1, Wolfram2}.

The asymptotic expansion of the elliptic integrals around $z = 1$
leads to:

\begin{equation}
\frac{\partial A\left( z \right) }{\partial z}= \rho - 12\ln 2 + 4 +
4\ln \left| {z - 1} \right|~,
\end{equation}
where $\rho = \int\limits_0^{2\pi } {\int\limits_0^1 {\frac{y\left(
{1 + \cos \varphi } \right)dyd\varphi }{\sqrt {1 + y^2 - 2y\cos
\varphi } }} } \simeq 5.6639$.

Integrating over $z$ we finally obtain:
\begin{eqnarray}
A\left( {1 + u} \right) - A\left( {1 - u} \right) \simeq 2\left(
{\rho - 12\ln 2 + 4\ln \left| u \right|} \right)u =\nonumber\\
\simeq8u\left( {\ln \left| {\xi u} \right|} \right)~,
\end{eqnarray}
with $\xi\simeq 0.51$.

\section{Evaluation of useful quadratures}
\label{Special}
We begin by calculating here the leading order term in the expansion
of the expression
\begin{equation}
I = \int\limits_0^{\arccos \frac{b}{2}} {\frac{\sqrt {\sin \varphi }
}{\cos \varphi }d\varphi }~~~ \rm{for}~ b \ll 1~,
\end{equation} .

Expanding the upper limit and setting $t = \tan \frac{\varphi }{2}$,
the integral becomes:

\begin{eqnarray}
I &\simeq& 2\sqrt 2 \int\limits_0^{1 - \frac{b}{2}} {\frac{dt}{1 -
t^2}\sqrt {\frac{t}{1 + t^2}} } \nonumber\\
&\simeq& \sqrt 2
\int\limits_0^{1 - \frac{b}{2}} {dt\sqrt {\frac{t}{1 + t^2}} \left(
{\frac{1}{1 - t} + \frac{1}{1 + t}} \right)}~,
\end{eqnarray}

The divergence in the limit $b \to 0$ stems only from the first term
and we therefore proceed to try isolating the singularity:

\begin{equation}
\label{R10}
\begin{array}{c}
 I \simeq \sqrt 2 \int\limits_0^{1 - \frac{b}{2}} {dt\sqrt {\frac{t}{1 +
t^2}} \frac{1}{1 - t}} + \sqrt 2 \int\limits_0^1 {dt\frac{1}{1 +
t}\sqrt
{\frac{t}{1 + t^2}} }+O(b)\simeq \\
 \simeq - \sqrt 2 \int\limits_0^{1 - \frac{b}{2}}
{dt\sqrt {\frac{t}{1 + t^2}} \left( {\ln \left( {1 - t} \right)}
\right)^\prime } + 0.5256+O(b) ~.\\
\end{array}
\end{equation}

An integration by parts is used in the remaining integral to further
isolate the singular contribution:

\begin{eqnarray}
I&\simeq& - \left. {\sqrt {\frac{2t}{1 + t^2}} \left( {\ln \left( {1
- t} \right)} \right)} \right|_0^{1 - \frac{b}{2}} \nonumber\\
&+& \sqrt 2 \int\limits_0^{1 - \frac{b}{2}} {dt\frac{d}{dt}\left(
{\sqrt {\frac{t}{1 + t^2}} } \right) \ln \left( {1 - t} \right)} +
0.5256 +O(b)~.\nonumber\\
\end{eqnarray}

At this point the non singular second term is  calculated
numerically so that the final result is:

\begin{equation}
\label{Rq1} I \simeq \ln b + 0.9475~.
\end{equation}

The following integral is used in the proof of the HF instability
theorem:

\begin{equation}
J = \int\limits_1^{\sqrt \varsigma } {\int\limits_1^{\sqrt \varsigma
} {\frac{dudv}{\sqrt {\left( {u^2 - v^2} \right)^2 +
\frac{u^2v^2b^2}{2}} }} }~,
\end{equation}
for $b \ll 1$.

Because the singular behavior originates from the region where
$u\simeq v$, in order to find the leading order term, we approximate
$u^2 - v^2$  with $2u(u - v)$. Since the function is symmetric with
respect to the interchange of $u$ and $v$ we can use the relation
$\int\limits_1^{\sqrt \varsigma } {du\int\limits_1^{\sqrt \varsigma
} {dv~f(u,v)  } } = 2\int\limits_1^{\sqrt \varsigma }
{du\int\limits_1^u {dv~f(u,v) } }$.

Then:
\begin{eqnarray}
J &\simeq& \int\limits_1^{\sqrt \varsigma }
{\frac{du}{u}\int\limits_1^u {\frac{dv}{\sqrt {\left( {u - v}
\right)^2 + \frac{u^2b^2}{8}} }} } \simeq -\int\limits_1^{\sqrt
\varsigma } {\frac{du}{u}\ln \left( {\frac{bu}{2\sqrt 2 }}
\right)}\nonumber\\
&+& \int\limits_1^{\sqrt \varsigma } {\frac{du}{u}\ln \left( {u - 1
+ \sqrt {\left( {u - 1} \right)^2 + \frac{u^2b^2}{8}} } \right)}~.
\end{eqnarray}

The integrand of the first term has a logarithmic singularity in the
limit of small $b$ while the second one is instead analytic in this
limit. The main contribution to the integral is therefore:
\begin{equation}
\label{Rq2} J \simeq - \int\limits_1^{\sqrt \varsigma }
{\frac{du}{u}\ln \left( {\frac{bu}{2\sqrt 2 }} \right)} \simeq - \ln
\varsigma \ln b~.
\end{equation}
\bibliography{Simion}

\begin{thebibliography}{20}
\bibitem[{com({\natexlab{a}})}]{comment_Hartree_term_cancellation}
\bibinfo{note}{Stability of a CDW state can only be achieved if the associated
  Hartree term is cancelled by a corresponding deformation of the neutralizing
  background. The simplest such case being represented by the case of the so
  called deformable jellium. A rather interesting case is discussed in
  \cite{GandO79}.}
\bibitem[{com({\natexlab{b}})}]{comment_HFsols}
\bibinfo{note}{That the states of Eqs.~(\ref{wf_nonint_p}) and
  (\ref{wf_nonint_m}) are solutions of the Hartree-Fock problem was shown in
  Ref.~\cite{ChesiGFG}.}
\bibitem[{com({\natexlab{c}})}]{comment_genChi1}
\bibinfo{note}{Notice that in this case, the generalized chirality, as defined
  for instance in Ref.\cite{ChesiGFG}, also equals one.}
\end{thebibliography}
\end{document}